\documentclass[aps,prl,twocolumn,showpacs]{revtex4}
\usepackage{amsbsy,latexsym}
\usepackage{amsfonts}
\usepackage[mathscr]{eucal}
\newcommand{\Fb}{\overline{F}}
\newcommand{\SU}{{\rm SU}}
\newcommand{\SO}{{\rm SO}}
\newcommand{\Ad}{{\rm Ad}}
\newcommand{\tr}{{\rm tr}\,}

\begin{document}
\title{Minimal measurements of the gate fidelity of a qudit map}
\author{E.~Bagan, M.~Baig and R.~Mu{\~n}oz-Tapia}
\affiliation{Grup de F{\'\i}sica Te{\`o}rica \& IFAE,
Facultat de Ci{\`e}ncies, Edifici Cn, Universitat Aut{\`o}noma de
Barcelona, 08193 Bellaterra (Barcelona) Spain}

\date{July 26, 2002}

\begin{abstract}
We obtain a simple formula for the average gate fidelity of a
linear map acting on qudits. It is given in terms of minimal sets
of pure state preparations alone, which may be interesting from
the experimental point of view. These preparations can be seen as
the outcomes of certain minimal positive operator valued measures. The
connection of our results with these generalized measurements
is briefly discussed.
\end{abstract}


\maketitle The interaction with the environment has a significant
negative impact on any physical implementation of a unitary gate
or a quantum channel. Decoherence, turning pure states into mixed
states, creeps up and unitarity is lost in the Hilbert subspace
of the signal states. The characterization of the quality of such
implementations is, hence, of utmost importance for quantum
computation from both experimental and theoretical point of
view~\cite{cirac}.

A physical quantum gate or channel is best described by a
trace-preserving linear map or superoperator $\cal E$, which is
assumed to be an approximation of a unitary operator $U$
characterizing the quantum gate ($U=\openone$ for a quantum channel).
The average gate fidelity
\begin{equation}
\Fb({\cal E},U)=\int d\psi \;\tr \left[U\rho_\psi U^\dagger {\cal E}(\rho_\psi)\right],
\label{F(E,U) integral}
\end{equation}
where  $d\psi$ is the invariant Haar measure on the space of pure
states $|\psi\rangle$ (i.e., $d\psi=d\psi'$ if
$|\psi'\rangle=U|\psi\rangle$) and
$\rho_\psi=|\psi\rangle\langle\psi|$, is recognized as a
convenient figure of merit and is widely used to assess the
quality of a quantum gate or channel.

For gates acting on qubits, Bowdrey {\em et al.}~\cite{bowdrey}
have recently derived a simple formula for $\Fb({\cal E},U)$, from
which they obtain a convenient way of actually measuring the gate
fidelity in a laboratory. It amounts to replacing the integral
in~(\ref{F(E,U) integral}) by a sum over a finite number of pure
states whose Bloch vectors point at the vertices of an octahedron
or a tetrahedron inscribed in the Bloch sphere.
 Nielsen~\cite{nielsen} has
obtained a similar formula for qudits (i.~e., quantum states
belonging to a $d$-dimensional Hilbert space) in terms of a
finite set of unitary operations $U_j$ orthogonal with respect to
the Hilbert-Schmidt inner product, namely, such that $\tr(
U_i^\dag U_j)=d\,\delta_{ij}$. His derivation is based on
Horodeckis' formula connecting $\Fb({\cal E},\openone)\equiv
\Fb({\cal E})$ with the entanglement fidelity~\cite{horodeckis},
for which he also provides a simple proof. Our aim is to give a
more natural generalization of Bowdrey's formula so that only
measurements over a (minimum) number of pure states need to be
performed to verify $\Fb$ experimentally. Interestingly enough,
the solution of this problem is not far removed from that of
obtaining minimal positive operator valued measurements (POVMs)
in the context of optimal communication of directions through a
quantum (spin) channel, which has received much attention over
the last few years~\cite{lots,product,lpt}.

In the first part of this letter we derive a general expression
of $\Fb$ in terms of the $\SU(d)$ group generators. In the second
part we  write $\Fb$ as an average of measurements  over a finite
and minimal number of pure state preparations, which may be
experimentally relevant.

We first notice that due to the invariance of $d\psi$ one has
$\Fb({\cal E},U)=\Fb({\cal E}',\openone)=\Fb({\cal E}')$, where ${\cal
E}'(\rho)\equiv{\cal E}(U^\dagger\rho U)$. Hence we only need to
consider the simpler form $\Fb({\cal E})$ without any loss of
generality. Furthermore, the uniform measure $d\psi$ can be
effectively realized in the following way: $ \int d\psi {\cal
F}(\psi)\equiv\int d U\, {\cal F}(U \psi_0) $, where ${\cal F}$
is any function of  $|\psi\rangle$, $|\psi_0\rangle$ is a fixed
reference state, and $d U$ is the Haar measure of $\SU(d)$
normalized such that $\int dU=1$. This ensures that $\int d\psi
{\cal F}(\psi)=\int d\psi {\cal F}(U'\psi)$ for any $U'\in
\SU(d)$. Obviously, not all the $d^2-1$ parameters involved in
$dU$ are physically significant. E.g., for qubits ($d=2$), $d\psi=
dn$ is the uniform measure on the 2-sphere $\mathbb{S}^2$, which
can be parametrized by the Euler angles $\alpha$ and $\beta$. If
$|\psi_0\rangle$ is an eigenstate of the Pauli matrix $\sigma_z$,
any function of $|\psi\rangle$ is independent of the third Euler
angle~$\gamma$. Hence, $\int d\psi\, {\cal F} =[\int d\gamma/(2
\pi)=1]\times \int dn \, {\cal F}= \int dU \, {\cal F}$.  For
qudits, $dn$ must be  replaced by the invariant measure of
$\SU(d)/[\SU(d-1)\times {\rm U}(1)]$, since any reference state
$|\psi_0\rangle$ is now invariant under $\SU(d-1)\times {\rm
U}(1)$. Note  that the number of parameters match, as qudits
depend  on  $2(d-1)$ real variables. With all this we finally have
\begin{equation}
\Fb({\cal E})=\int dU\,\tr\left[ U\rho_0 U^\dagger {\cal
E}(U\rho_0 U^\dagger) \right], \label{Fb=int dU}
\end{equation}
where we have defined $\rho_0\equiv|\psi_0\rangle\langle\psi_0|$.

There exists a $(d^2-1)$-dimensional unitary vector with components
$n_0^a$, $a=1,2,\dots d^2-1$,  such that the density matrix $\rho_0$
can be written as
\begin{equation}
\rho_0={\openone\over d} + k_d n_0^a T_a\equiv {\openone\over d} +
k_d \vec n_0\cdot\vec T,
\end{equation}
where $k_d=\sqrt{2(d-1)/d}$ and $\{T_a\}$ are the (hermitian and
traceless) generators of $\SU(d)$, normalized so that
$\tr(T_aT_b)=\delta_{ab}/2$. They can be chosen as
$T_a=\lambda_a/2$, where $\lambda_a$ are a generalization of the
Gell-Mann matrices of $\SU(3)$. Throughout this letter a sum over
repeated latin indexes is understood. It is straightforward to
obtain
\begin{equation}
\vec n_0={2\over k_d}\tr(\rho_0 \vec T).
 \label{vec n_0}
\end{equation}
We now recall the well known relation
\begin{equation}
UT_aU^\dagger=(\Ad U)_a{}^b T_b,
\end{equation}
where $\Ad$ stands for the adjoint representation of $\SU(d)$. We
also recall the orthogonality of the irreducible representations
of compact groups, which in the present case implies
\begin{eqnarray}
\int dU\, (\Ad U)_a{}^b&=&0, \label{orth rel-1}\\
\int dU\, (\Ad U)_a{}^b (\Ad
U)_c{}^d&=&{\delta_{ac}\delta^{bd}\over d^2-1}. \label{orth rel-2}
\end{eqnarray}
 Using that $\cal E$ is linear and trace-preserving one gets
\begin{equation}
\Fb({\cal E})={1\over d}+{2\over
d(d+1)}\sum_{a=1}^{d^2-1}\tr\left[T_a{\cal E}(T_a)\right].
\label{generalized Bowdrey}
\end{equation}
This is the generalization of Bowdrey's formula
\begin{equation}
\Fb({\cal E})={1\over 2}+{1\over 3}
\sum_{i=x,y,z}\tr\left[{\sigma_i\over 2}{\cal E}\left({\sigma_i\over 2}\right)\right]
\end{equation}
for qubits. That concludes the first part of this letter.

The key ingredient of the derivation above is the orthogonality
relation of the group representations, exemplified by
Eqs.~\ref{orth rel-1} and~\ref{orth rel-2}. We will show that it
is possible to find a discrete version of these equations.
Namely, one can find a {\em finite} set of $\SU(d)$ elements
$\{U_r\}$ and positive constants $\{c_r\}$ such that
\begin{eqnarray}
\sum_r  c_r (\Ad U_r)_a{}^b&=&0
\label{orth rel finite-1}\\
\sum_r c_r  (\Ad U_r)_a{}^b (\Ad
U_r)_c{}^d&=&{\delta_{ac}\delta^{bd}\over d^2-1}. \label{orth rel
finite-2}
\end{eqnarray}
With this we can reverse the steps going from~(\ref{Fb=int dU})
to~(\ref{generalized Bowdrey}) using the relations~(\ref{orth rel
finite-1}) and (\ref{orth rel finite-2}) instead of their
continuous version~(\ref{orth rel-1}) and (\ref{orth rel-2}), and
obtain
\begin{equation}
\Fb({\cal E})=\sum_{r} c_r \tr\left[\rho_r {\cal E}(\rho_r
)\right]. \label{Fb-general}
\end{equation}
This equation has a very convenient form which allows setting up
experimental tests to determine the fidelity of a gate or
channel, as will be discussed below.

Let us briefly discuss the solutions to Eqs.~\ref{orth rel
finite-1} and~\ref{orth rel finite-2}.   The idea is to
generalize the concept of a finite set of {\em isotropically
distributed} unit vectors introduced in~\cite{product} and adapt
it to the problem at hand. A sufficient condition for
(\ref{Fb-general}) to hold can be obtained by
contracting~(\ref{orth rel finite-1}) and~(\ref{orth rel
finite-2}) with $n_0^a$. If we define
\begin{equation}
n_0^a (\Ad U_r)_a{}^b\equiv n_r^b,
\label{condition-2}
\end{equation}
conditions~(\ref{orth rel finite-1}) and~(\ref{orth rel finite-2}) lead  to
\begin{equation}
\sum_r  c_r n_r^b=0;\quad \sum_r c_r
n_r^bn_r^d={\delta^{bd}\over d^2-1}.
 \label{orth vectors}
\end{equation}
In this sense, we may qualify the set $\{n_r\}$  as isotropically
distributed  (as far as the adjoint representation $\SU(d)$ is
concerned). We have traded the problem of finding $\{U_r\}$ for
that of finding $\{n_r\}$.  Note, however that the set of
matrices $\{\Ad U \}$ is a proper subgroup of $\SO(d^2-1)$.
Hence, not any vector on the sphere $\mathbb S^{d^2-2}$ is
admissible.  We will come back to this issue below. The
relations~(\ref{orth rel finite-1}), (\ref{orth rel finite-2})
and~(\ref{orth vectors}) also appear in the rather different
context: the construction of finite positive operator valued
measurements which are optimal for communicating a
direction~\cite{product,lpt,alp}. Leaving aside an overall trivial
normalization (from Eq.~\ref{orth vectors} it follows that $\sum_r
c_r=1$ whereas in~\cite{alp} $\sum_r c_r=d$), the results in those
papers can be readily  used here. In particular, it is proved
in~\cite{alp} that solutions of~(\ref{orth vectors}) exist and the
minimal one is given by a set of $d^2$ vectors pointing at the
vertices of a regular hypertetrahedron or, more properly,
$(d^2-1)$-simplex, inscribed on $\mathbb S^{d^2-2}$. This
hypertetrahedron is defined by the condition
\begin{equation}
\vec n_r\cdot\vec n_s=-{1\over d^2-1},\quad r\not=s,
\label{n_r.n_s}
\end{equation}
 and the exact overall orientation has to be
chosen so that all vectors $\vec n_r$ are of the
form~(\ref{condition-2}).  For this hypertetrahedron all the
coefficients $c_r$ are equal: in our notation $c_r=1/d^2$,
$r=1,2,\dots,d^2$. A explicit form of $\vec n_r$ for $\SU(3)$ can
be found in~\cite{alp}, where also the general case is briefly
discussed. The solution is more conveniently expressed in terms
of states $|\psi_r\rangle$ such that
\begin{equation}
|\psi_r\rangle\langle\psi_r|\equiv \rho_r=U_r\rho_0 U_r^\dagger ={\openone\over d}+k_d\vec n_r\cdot
\vec T . \label{rho_r}
\end{equation}
Then~(\ref{n_r.n_s}) translates into
\begin{equation}
\left|\langle\psi_r|\psi_s\rangle\right|^2={1\over d+1},\quad
r\not=s. \label{psi_r.psi_s}
\end{equation}
Since this equation is a condition on states,
Eq~\ref{condition-2} is automatically satisfied by the
corresponding Bloch vectors.

For $\SU(3)$ a solution of~(\ref{psi_r.psi_s}) has the simple form
\begin{equation}
|\psi_r\rangle={1\over\sqrt{2}}\pmatrix{1\cr{\rm e}^{2(r-1)\pi
i/3}\cr0},\qquad r=1,2,3. \label{psi_r su(3)}
\end{equation}
The remaining six states are obtained by applying cyclic
permutations to the components of
$|\psi_{\mbox{\scriptsize$1$--$3$}}\rangle$ (a different choice
of states is given in~\cite{alp}). From a experimental point
of view, the states (\ref{psi_r su(3)}) have a very appealing
form; each of them involves a linear combination of \emph{only
two} states of the computational basis. If the  qutrit is
implemented by say three atomic levels, only two levels need to be
manipulated to prepare each one of the $\rho_r$.

In the $\SU(2)$ case, a solution in terms of $\vec n_r$ is more
simple and transparent, mainly because  $\SU(2)$ is isomorphic to
$\SO(3)$,  which makes~(\ref{condition-2}) trivially
satisfied by any vector of~$\mathbb S^2$. A compact solution is
given by
\begin{equation}
\vec n_1={1\over\sqrt3}(1,1,1),\quad \vec n_2={1\over\sqrt3}(-1,-1,1),
\end{equation}
\label{n_r su(2)}
and $\vec n_3$, $\vec n_4$ are again obtained by applying
cyclic permutations to the components of $\vec n_2$.

For the hypertetrahedra discussed above, Eq.~\ref{Fb-general} can
be cast as
\begin{equation}
\Fb({\cal E})={1\over d^2}\sum_{r=1}^{d^2} \tr\left[\rho_r {\cal E}(\rho_r )\right],
\label{Fb in terms of rho_r}
\end{equation}
which is the $\SU(d)$ generalization of Bowdrey's $\SU(2)$
formula. This equation is our main result and provides a
remarkably simple procedure for measuring $\Fb({\cal E})$. One
just has to average the fidelities for $d^2$ isotropically
distributed pure states. Notice that all state preparations
$\rho_r$ have the same weight in the average, thus reducing
systematic errors. In some sense,  Eq. \ref{Fb in terms of rho_r}
can be regarded as the average survival rate of the states
$\{\rho_r\}$ in  a quantum channel characterized by the linear
map $\cal E$. There is a alternative way of writing~(\ref{Fb in
terms of rho_r}) which may provide further insight. It is
straightforward to verify that the projectors $O_r\equiv \rho_r/d$
are the complete set of positive operators (i.e., $\sum
O_r=\openone$) of a minimal POVM. Thus, the states $\rho_r$ are
just the preparations produced by the device characterized by
$\{O_r\}$. We can write~(\ref{Fb in terms of rho_r}) as
\begin{equation}
\Fb({\cal E})={1\over d}\sum_{r=1}^{d^2} \tr\left[O_r {\cal
E}(\rho_r )\right] \label{Fb in terms of O_r}.
\end{equation}
We readily see that the same device $\{O_r\}$ could be used for
the preparations of the pure states $\{\rho_r\}$ as well as for
the measurements over $\{{\cal E}(\rho_r)\}$.

Although we have presented the results for the minimal sets of
$\rho_r$, for practical reasons, one may wish to use a larger
number of states. This possibility is easily implemented in this
framework, as conditions~(\ref{condition-2}) and~(\ref{orth
vectors}) are not specific of minimal sets but entirely  general.
E.g., for qubits one can find a set of six states whose Bloch
vectors point at the vertices of a regular octahedron
\cite{lpt,lots,bowdrey}). In this case there is a simple setting
for preparing the states, just three Stern-Gerlach's oriented
along the three orthogonal directions.

We are grateful to A. Ac{\'\i}n and A. Bramon for helpful
conversations. Financial support from contracts CICYT AEN99-0766
and CIRIT 2000SGR-00063 is acknowledged.

\end{document}